\documentclass[journal,twoside,web]{ieeecolor}

\usepackage{generic}
\usepackage{cite}
\usepackage{amsmath,amssymb,amsfonts}
\usepackage{algorithmic}
\usepackage{graphicx,multicol}
\usepackage{textcomp}
\def\BibTeX{{\rm B\kern-.05em{\sc i\kern-.025em b}\kern-.08em
    T\kern-.1667em\lower.7ex\hbox{E}\kern-.125emX}}

\usepackage{mathbbol}
\usepackage{wrapfig}
\usepackage{mathrsfs}
\usepackage{url}
\usepackage{array}
\usepackage{amsmath,amsfonts, amssymb}

\usepackage{generic}
\usepackage{cite}
\usepackage{amsmath,amssymb,amsfonts}
\usepackage{algorithmic}
\usepackage{graphicx}
\usepackage{algorithm,algorithmic}
\usepackage{hyperref}
\usepackage{textcomp}

\newtheorem{theorem}{\textbf{Theorem}}

\newtheorem{definition}{\textbf{Definition}}
\newtheorem{assumption}{\textbf{Assumption}}
\newtheorem{proposition}{\textbf{Proposition}}

\newtheorem{remark}{Remark}

\usepackage{graphicx}
\usepackage{tabularx}
\usepackage{array}
\usepackage{epstopdf}
\usepackage{lipsum}
\usepackage{tablefootnote}
\usepackage{subfigure}

\newcommand{\rev}[1]{{#1}}

\newcommand{\St}{\mathtt{S}}
\newcommand{\yint}{y^*_{\mathtt{INT}}}
\newcommand{\zsint}{z^*_{\bar{\mathtt{S}},\mathtt{INT}}}
\newcommand{\It}{\mathtt{I}}
\newcommand{\Stb}{\bar{\mathtt{S}}}
\newcommand{\Itb}{\bar{\mathtt{I}}}

\newcommand{\zsdag}{z^\dagger_{\bar{\mathtt{S}}}}
\newcommand{\zidag}{z^\dagger_{\bar{\mathtt{I}}}}
\newcommand{\zsbar}{z^\star_{\bar{\mathtt{S}}}}
\newcommand{\zibar}{z^\star_{\bar{\mathtt{I}}}}

\newcommand{\zp}{\mathbf{z}}

\newcommand{\yee}{y_{\mathtt{EE}}}

\newcommand{\Pt}{\mathtt{P}}
\newcommand{\Ut}{\mathtt{U}}

\newcommand{\Pb}{\mathbb{P}}

\newcommand{\betap}{\beta_{\mathtt{P}}}
\newcommand{\betau}{\beta_{\mathtt{U}}}

\newcommand{\ignore}[1]{}
\emergencystretch=\maxdimen
\allowdisplaybreaks

\begin{document}
\title{\LARGE \bf
Bayesian Persuasion for Containing SIS Epidemics with Asymptomatic Infection
\author{Ashish R. Hota, \IEEEmembership{Senior Member, IEEE}, Abhisek Satapathi, and Urmee Maitra, \IEEEmembership{Graduate Student Member, IEEE}
\thanks{The authors are with the Department of Electrical Engineering, Indian Institute of Technology (IIT) Kharagpur, Kharagpur, West Bengal, India, 721302. (e-mail: ahota@ee.iitkgp.ac.in, abhisek.ee@iitkgp.ac.in, urmeemaitra93@kgpian.iitkgp.ac.in).}}
}

\maketitle
\thispagestyle{empty}


\begin{abstract}
We investigate the strategic behavior of a large population of agents who decide whether to adopt a costly partially effective protection or remain unprotected against the susceptible-infected-susceptible epidemic. In contrast with most prior works on epidemic games, we assume that the agents are not aware of their true infection status while making decisions. We adopt the Bayesian persuasion framework where the agents receive a noisy signal regarding their true infection status, and maximize their expected utility computed using the posterior probability of being infected conditioned on the received signal. We characterize the stationary Nash equilibrium of this setting \rev{under suitable assumptions}, and identify conditions under which partial information disclosure leads to a smaller proportion of infected individuals at the equilibrium compared to full information disclosure, and vice versa.
\end{abstract}

\section{Introduction}

Effective containment of spreading processes is an important problem in many settings, such as to control the spread of infectious diseases in society, malware and virus in computer networks, and opinions in social networks \cite{nowzari2015optimal}. Centralized approaches to contain the spread of epidemics, based on optimal and predictive control strategies, are often not scalable in settings involving a large population of agents \cite{nowzari2015optimal,zino2021analysis}. In fact, individuals largely decide whether to adopt protection (e.g., use masks against an infectious disease) or not, in a strategic and decentralized manner depending on the risk and incentives that they encounter. 

Consequently, past works have explored the effectiveness of game-theoretic decision-making in reducing the impact of epidemics in a variety of settings, including but not limited to time-varying and networked interactions \cite{hota2022impacts,eksin2016disease}, adoption of masks or social distancing \cite{satapathi2023coupled,paarporn2023sis} and vaccines \cite{hota2023learning}, and diseases with asymptomatic infections \cite{hota2023learning,olmez2022modeling}; see \cite{huang2022game} for a recent review. Similarly, recent works, such as \cite{martins2023epidemic}, have borrowed tools from dynamical systems theory to design control strategies to influence the behavior of agents in such game-theoretic settings. However, most of the above works have assumed that the agents are aware of their true infection status. 

In contrast, it was observed during the COVID-19 pandemic that asymptomatic infections pose a severe challenge in containing the spread of epidemics \cite{bi2021role}. Since it is not practical to frequently test an entire population to detect the disease, a large number of smartphone applications have emerged that provide personalized infection risk status to individuals by monitoring their interaction pattern and proximity to other infected individuals \cite{akinbi2021contact,lewis2021contact,google_app}, and induce them to adopt protection to reduce risk of becoming infected. However, the recommendations provided by such applications could be potentially erroneous, and may even be biased by design to try and elicit a certain type of response from the population. In addition, individuals often do not blindly follow the recommendations of such applications, rather evaluate the risk of becoming infected and cost of adopting protection in light of the information received from the application as well as other sources.  

Therefore, it is essential to analyze the impact of such information disclosure on protection adoption behavior by a large population of selfish agents against stealthy spreading processes. Despite its significance, this problem has not been rigorously investigated in the past. In this paper, we investigate this problem in the framework of {\it Bayesian persuasion} \cite{kamenica2011bayesian} and population games \cite{sandholm2010population}, and explore when it is possible to reduce infection prevalence by appropriate design of signalling schemes. 

In the Bayesian persuasion framework, introduced in \cite{kamenica2011bayesian}, the principal sends a noisy signal to the agents regarding the state of a variable which the agents cannot observe. The agents choose an action which maximizes their expected utility computed using the posterior distribution of the state conditioned on the received signal. This framework is often viewed as an alternative to classical mechanism design since the incentive provided to the agents is informational rather than monetary. In addition to being widely applied in economics \cite{kamenica2019bayesian,bergemann2019information}, this framework has also been investigated in many engineering applications, including managing congestion in transportation networks \cite{ferguson2022avoiding,verbree2021inferring,wu2021value,acemoglu2018informational}, internet routing with incomplete knowledge \cite{seaton2023competitive}, and security games \cite{farhadi2018static}. 

In a recent paper \cite{pathak2022scalable}, the authors examine a networked public goods game setting inspired by epidemic risks, where the principal sends a signal regarding disease severity to the agents. The agents then decide their level of contribution to the public good. The main result shows that design of the signalling scheme to induce greater contribution to the public good is independent of the structure of the network under suitable assumptions. However, the authors do not consider any specific epidemic model to capture the fact that epidemic severity or prevalence is often time-varying and evolves as a function of the actions by the agents. Furthermore, apart from \cite{pathak2022scalable}, we are not aware of any other work that has examined the problem of containing epidemics via Bayesian persuasion. 

In order to lay the foundations towards understanding the impact of strategic information disclosure in shaping human behavior against epidemics, we consider the susceptible-infected-susceptible (SIS) epidemic setting with a large population of homogeneous individuals or agents. Each individual decides whether or not to adopt protection, which comes with a cost and \rev{reduces its likelihood of becoming infected or transmitting infection to others}. An agent is not aware of its true infection state, but receives a signal regarding its infection status from the principal. It then computes the posterior probability of being infected by assuming the overall proportion of agents who are infected as prior, and computes the expected utility under both available actions.

We define the \rev{population state to be the tuple consisting of the proportion of agents who are infected, and the proportions of agents who remain unprotected when they receive susceptible and infected signals, respectively. A population state is said to be a stationary Nash equilibrium (SNE) when no individual agent can improve its expected reward by choosing a different action, and the infected proportion is at the endemic equilibrium for the given protection adoption strategies of the agents.} We characterize the stationary Nash equilibrium of this game under \rev{suitable} assumptions, and determine the range of parameters under which partial information disclosure (PID) leads to a smaller fraction of infected agents at the equilibrium compared to truthful or full information disclosure (FID). \rev{We further show that when the information revealed to the agents is substantially different from the truth}, it may lead to a worse outcome compared to FID. Our numerical results show the convergence of the coupled epidemic and evolutionary learning dynamics, captured by Smith dynamics \cite{sandholm2010population}, to the SNE, and provide insights into how infected proportion at the SNE can be reduced via suitable information disclosure. 

\section{Problem Formulation}\label{sec:strategic_protection}

\subsection{Infection States and Signalling Scheme} 

We consider an SIS epidemic model over a large population of homogeneous individuals or agents. Each agent belongs to one of the two states: \emph{susceptible} $\St$ or \emph{infected} $\It$. Proportions of infected and susceptible agents in the population at time $t$ are denoted by $y(t)$ and $1 - y(t)$, respectively, with $y(t) \in [0,1]$. We assume that the infections are asymptomatic. Thus, an individual is not aware of its exact infection status. We assume the presence of a signalling scheme that gives noisy information to the agents regarding their infection status. For instance, this can be achieved by a smartphone application that keeps tracks of the agents' social interactions, and computes their likelihood of being infected. However, the signalling scheme is not free of error, either due to inaccuracy or by design. 

We denote the signal received by an agent by $x \in \{\Stb,\Itb\}$. An agent receiving a signal $\Stb$ (respectively, $\Itb$) does not necessarily imply that its true state is susceptible ($\St$) (respectively, $\It$). The probability of a susceptible agent receiving a signal $\Stb$ is denoted by $\mu_{\St} \in [0,1]$. Formally, 
$$\mu_{\St} := \Pb[x = \Stb \mid \St], \quad \mu_{\It} := \Pb[x = \Itb \mid \It] \in [0,1].$$ Similarly, $\Pb[x = \Itb \mid \St] = 1-\mu_{\St}$ and $\Pb[x = \Stb \mid \It] = 1-\mu_{\It}$. When $\mu_{\St}=1$, then, a susceptible agent always receives its true infection status via the signal. The regime where $\mu_{\St} = \mu_{\It} = 1$ is referred to as the {\it full information disclosure} (FID) regime as the signal reveals the true infection status.

At any given time $t$, we assume that each agent is aware of the proportion of infected nodes in the population $y(t)$. \rev{This assumption is motivated by the observation during COVID-19 that a small fraction of the population (about $0.5-2\%$) used to undergo testing every day \cite{daily_test_data}, the corresponding test positivity rate data was made available to the public, and known to individuals who did not undergo testing. Since the number of tests administered were of the order of hundreds of thousands, individuals who underwent testing formed a representative population, and the positivity rate among the tested population would be a good benchmark for the infected fraction in the entire population.}

Thus, before it receives the signal, the agent assumes that it is infected with probability $y(t)$ (which acts as the prior). Once it receives the signal $x$, it computes its (posterior) {\it belief} regarding its true infection state in a Bayesian manner as  
\begin{align}
    \pi^+[\It | x] &= \frac{\mathbb{P}[x | \It] y(t)}{\mathbb{P}[x | \It] y(t) + \mathbb{P}[x | \St] (1-y(t))}.
\end{align}
In other words, the belief is the posterior probability of being infected given the signal it received, with the prior infection probability being $y(t)$. Specifically, we have
\begin{align*}
    \pi^+[\St | \Stb] &= \frac{\mu_{\St}(1 - y)}{(1 - \mu_{\It})y + \mu_{\St}(1 - y)},
    \\ \pi^+[\St | \Itb] &= \frac{(1 - \mu_{\St})(1 - y)}{\mu_{\It} y + (1 - \mu_{\St})(1 - y)}.
\end{align*}
It naturally follows that $\pi^+[\It | x] = 1-\pi^+[\St | x]$. 

\subsection{Epidemic Evolution under Protection Adoption} 
 
Two actions are available to the agents in the form of adopting protection $\Pt$, or remaining unprotected $\Ut$; denoted formally as $a \in \{\Pt,\Ut\}$. A susceptible agent adopting protection \rev{reduces its likelihood of becoming infected by a factor $\alpha \in (0, 1)$} compared to a susceptible agent which remains unprotected. An infected agent that adopts protection (or remains unprotected) spreads infection with \rev{rate} $\beta_{\Pt}$ ($\beta_{\Ut}$,  respectively), with $0 < \beta_{\Pt} < \beta_{\Ut}$. Since the agents are not aware of their true infection states, they make their decisions as a function of the signal they receive. Consequently, we let $z_{\Stb}(t)$ (respectively, $z_{\Itb}(t)$) denote the proportions of agents who remain unprotected among the agents that receive signal $\Stb$ (respectively, $\Itb$) at time $t$. 

However, the evolution of infection prevalence depends on the proportions of susceptible and infected agents that remain unprotected, denoted by $z_{\St}(t)$ and $z_{\It}(t)$, respectively. For a given true infection state $r \in \{\St,\It\}$, we have
\begin{align*}
    \mathbb{P}[\Ut | r] = z_{r}  = \mathbb{P}[\Ut | \Stb] \mathbb{P}[\Stb | r] + \mathbb{P}[\Ut | \Itb] \mathbb{P}[\Itb | r]
\end{align*}
following the law of total probability. Thus,
\begin{align}
    z_{\St} &= z_{\Stb} \mu_{\St} + z_{\Itb} (1 - \mu_{\St}),
    \label{eq:zs}
    \\z_{\It} &= z_{\Itb} \mu_{\It} + z_{\Stb} (1 - \mu_{\It}),
    \label{eq:zi}
\end{align}
where the dependence on time $t$ is suppressed for brevity of notation. Naturally, we have $z_{\St}, z_{\It}, z_{\Stb}, z_{\Itb} \in [0, 1]$. 

Following \cite{satapathi2023coupled}, the proportion of infected individuals evolves according to the SIS epidemic model with protection adoption as
\begin{align}
    \dot{y} &= \big[(1 \!- y) \rev{(\alpha (1 \!- z_{\St}) + z_{\St}) (\beta_{\Pt} (1 - z_{\It}) + \beta_{\Ut} z_{\It})}  -\! \gamma\big] y, \nonumber
    \\ & =: \big[(1 - y) \beta_{\mathtt{eff}}(z_{\Stb},z_{\Itb}) - \gamma\big]y,
    \label{eq:sis_dynamics}
\end{align}
where \rev{$\gamma > 0$ is the rate} with which an infected agent recovers, and $\beta_{\mathtt{eff}}(z_{\Stb},z_{\Itb})$ is the effective infection rate.

\subsection{Rewards and Utilities}

We now introduce the reward and utility that an agent incurs as a function of its state and chosen action as well as the actions chosen by other agents given by $z_{\Stb}, z_{\Itb}$ and the epidemic prevalence $y$. \rev{The tuple $\zp := (y,z_{\Stb}, z_{\Itb})$ is referred to as the {\it population state}.} We model our agents to be \emph{myopic}, i.e., the agents choose their actions based on the instantaneous rewards. \rev{Let $F[r, a; \textbf{z}]$ denote the instantaneous reward of an agent which belong to infection state $r$, chooses action $a$, given that the population state is $\textbf{z}$.}  

First, we tackle the case of an infected agent. We assume that an infected agent receives a reward of $-C_{\Pt}$ if it adopts protection, and a reward $-C_{\Ut}$ if it remains unprotected; here $C_{\Pt} > 0$ is the cost of adopting protection and $C_{\Ut} > 0$ represents the \rev{expected} penalty imposed by authorities on infected agents if they are found to violate isolation norms.\footnote{\rev{$C_{\Ut}$ can be interpreted as the actual penalty multiplied by probability of being found to violate isolation norms.}} Formally, we have
\begin{align*}
F[\It, \Ut;\zp] &= -C_{\Ut}, \qquad F[\It, \Pt;\zp] = -C_{\Pt}.
\end{align*}

\rev{If the agent is susceptible, then in addition to incurring the cost of protection $C_{\mathtt{P}}$, it also encounters the risk of becoming infected. The expected loss of becoming infected is the product of loss upon infection $L$, the instantaneous probability of becoming infected $(z_{\mathtt{I}} \beta_{\mathtt{U}} + (1-z_\mathtt{I})\beta_{\mathtt{P}})y$, and multiplying factor $\alpha$ if the agent opts for adopting protection. Formally,
\begin{align*}
    F[\St, \Ut;\zp] &= - L (z_{\It} \beta_{\Ut} + (1 - z_{\It}) \beta_{\Pt}) y, 
    \\&= -L ((\beta_{\Ut} z_{\Stb} + \beta_{\Pt} (1 - z_{\Stb})) (1 - \mu_{\It}) 
    \\& \qquad + (\beta_{\Ut} z_{\Itb} + \beta_{\Pt} (1 - z_{\Itb})) \mu_{\It}) y,
    \\ F[\St, \Pt;\zp] &= -C_{\Pt} + \alpha F[\St, \Ut;\zp].
\end{align*}
Since an infected agent is already infected, there is no immediate risk of becoming infected yet again. Hence, the second term in $F[\St, \Pt;\zp]$ is not present in $F[\It, \Pt;\zp]$.} In absence of knowledge of its true state, an agent computes its utility as the \emph{expected payoff} (dictated by its posterior belief regarding its infection state conditioned on the received signal) for each action, and chooses the action that gives a larger utility. We define the utility for an agent receiving signal $x \in \{\Stb,\Itb\}$ and choosing action $a \in \{ \Pt,\Ut\}$ as 
$$ U[x, a] := \pi^+[\St | x] F[\St, a] + \pi^+[\It | x] F[\It, a];$$
the dependence on $\zp$ is omitted for better readability. 


\section{Equilibrium Characterization}

\label{sec:eq_char}

In this section, we analyze the Nash equilibrium of the above setting under the following assumptions.

\begin{assumption}\label{assumption:main}
We assume the following.
\begin{itemize}
    \item The signal revealed to infected agents is truthful, i.e., $\mu_{\It} = 1$.
    \item $C_{\Pt} < C_{\Ut}$, which incentivizes infected agents to adopt protection.  
    \item Recovery rate $\gamma < \alpha \beta_{\Pt}$. 
\end{itemize}
\end{assumption}

\rev{Note that it is expected that a signalling scheme is biased towards one of the outcomes, i.e., if it has a considerable false alarm rate, then it should have a smaller rate of missed detection, and vice versa. We have assumed $\mu_{\mathtt{I}} = 1$ and $\mu_{\mathtt{S}} < 1$, motivated by the fact that in the context of an actual pandemic, such a scheme would prefer to err on the side of being safe. The parameter $C_{\mathtt{U}}$ is often set by the authorities to prevent infected individuals from not adopting protection, and hence, $C_{\mathtt{P}} < C_{\mathtt{U}}$ is assumed. Another motivation is that infected individuals may adopt protection due to reasons such as altruism (e.g., towards family members), and this is captured by assuming $C_{\mathtt{P}} < C_{\mathtt{U}}$. Further discussion on our assumptions are presented in Remark \ref{remark:assumption} at the end of this section.} 

In order to analyze the Nash equilibria of this setting, we first compute the difference in utilities obtained for each action. Observe that $\mu_{\It} = 1$ implies  
\begin{equation}\label{eq:pi_plus_s_ibar}
\pi^+[\St | \Stb] = 1, \quad \pi^+[\St | \Itb] = \frac{(1-y)(1-\mu_{\St})}{(1-y)(1-\mu_{\St}) + y}.
\end{equation}

For agents who receive signal $\Itb$, we compute 
\begin{align}
    U[\Itb, \Pt] &= \pi^+[\St | \Itb] F[\St, \Pt] + (1 - \pi^+[\St | \Itb]) F[\It, \Pt]\nonumber
    \\&= \alpha \pi^+[\St | \Itb] F[\St, \Ut] - C_{\Pt}, \nonumber\\
    U[\Itb, \Ut] &=  \pi^+[\St | \Itb] F[\St, \Ut] + (1 - \pi^+[\St | \Itb]) F[\It,\Ut]\nonumber
    \\&= \pi^+[\St | \Itb] F[\St, \Ut] - (1 - \pi^+[\St | \Itb]) C_{\Ut}, \nonumber
\\ \Rightarrow \Delta U[\Itb] & = U[\Itb, \Pt] - U[\Itb, \Ut] \label{eq:del_Ibar} 
    \\ & = -\pi^+[\St | \Itb] ((1-\alpha) F[\St, \Ut] + C_{\Ut}) + C_{\Ut} - C_{\Pt}. \nonumber
\end{align}

Similarly, for agents who receive signal $\Stb$, we compute 
\begin{align}
    U[\Stb, \Pt] &=  \pi^+[\St| \Stb] F[\St, \Pt] + (1 - \pi^+[\St| \Stb]) F[\It, \Pt]\nonumber
    \\&= \alpha F[\St, \Ut] - C_{\Pt} \nonumber
    \\U[\Stb, \Ut] &=  \pi^+[\St| \Stb] F[\St, \Ut] + (1 - \pi^+[\St| \Stb]) F[\It, \Ut] = F[\St, \Ut], \nonumber
\\ \Rightarrow \Delta U[\Stb] & = U[\Stb, \Pt] - U[\Stb, \Ut] \nonumber
    \\ & = -(1-\alpha) F[\St, \Ut]- C_{\Pt} \nonumber
    \\ & = (1-\alpha) L (\beta_{\Pt} + (\beta_{\Ut} - \beta_{\Pt}) z_{\Itb}) y- C_{\Pt}.
    \label{eq:del_Sbar}
\end{align}

Note that the above difference in utilities depend on the population state $\zp$, which is suppressed for better readability. We now formally define the notion of {\it stationary Nash equilibrium} for this setting.

\begin{definition}\label{def:SNE}
    The necessary and sufficient conditions for a population state $\zp^\star = (y^\star,z^\star_{\Stb},z^\star_{\Itb})$ to be a stationary Nash equilibrium (SNE) are given below:
    \begin{itemize}
    \item $y^\star = \yee(z^\star_{\Stb},z^\star_{\Itb}) := 1 - \frac{\gamma}{\beta_{\mathtt{eff}}(z^\star_{\Stb},z^\star_{\Itb})}$,
    \item for $x \in \{ \Stb,\Itb\}$, we have the following necessary conditions:
    \begin{itemize}
        \item $z^\star_{x} = 0 \Rightarrow \Delta U[x](\zp^\star) \geq 0$,
        \item $z^\star_{x} = 1 \Rightarrow \Delta U[x](\zp^\star) \leq 0$, 
        \item $z^\star_{x} \in (0,1) \Rightarrow \Delta U[x](\zp^\star) = 0$, and
    \end{itemize}
    \item for $x \in \{ \Stb,\Itb\}$, we have the following sufficient conditions:
    \begin{itemize}
        \item $\Delta U[x](\zp^\star) > 0 \Rightarrow z^\star_{x} = 0$,
        \item $\Delta U[x](\zp^\star) < 0 \Rightarrow z^\star_{x} = 1$,
        \item $\Delta U[x](\zp^\star) = 0 \Rightarrow z^\star_{x} \in [0,1]$.
    \end{itemize}
    \end{itemize}
\end{definition}
The term $\yee(z^\star_{\Stb},z^\star_{\Itb})$ denotes the infected proportion at the endemic equilibrium of the disease dynamics \eqref{eq:sis_dynamics} as a function of the proportions of agents not adopting protection. \rev{Our assumption $\gamma < \alpha \betap$ guarantees the existence of a unique nonzero endemic equilibrium for $z^\star_{\Stb},z^\star_{\Itb} \in [0,1]$.}

\begin{remark}
\rev{We have formulated the problem in the framework of {\it population games} \cite{sandholm2010population} which assumes that the number of agents is large (approaching infinity), and an individual agent changing its strategy does not alter the population state. Further, the payoffs as well as the equilibrium is defined in terms of the population state. An individual agent computes the payoff it would obtain for each available action at the current population state, and prefers to opt for the action which gives it the highest payoff. The SNE is defined as the population state at which no agent can choose a different action and obtain a strictly larger payoff. If at a SNE, nonzero fractions of individuals have opted for multiple actions, then the payoffs obtained by opting for each of those actions must be identical. Another way of interpreting the equilibrium is that each agent follows a mixed strategy to choose its action, with probability of picking an action given by the fraction of individuals opting for that action at the SNE population state. Consequently, following the law of large numbers, the fraction of individuals opting for each action ends up coinciding with the SNE population state. 

Extensive past research have shown that simple learning rules followed by a finite number of agents to revise their actions can be approximated in the (mean-field) limit as differential equations stated in terms of the population state or fractions opting for different actions \cite{sandholm2010population}. Since the framework of population games and evolutionary learning is fairly well established by now, most works directly analyze the equilibria and dynamics stated in terms of the population state instead of delving into individual strategy updates, which is also the convention we have adopted. One specific manner in which the present setting differs from the classical population game framework is that the payoffs depend on the fraction of infected agents $y(t)$ which is a time-varying quantity, and evolves as a function of the strategies ($z_{\bar{{\mathtt{S}}}}$ and $z_{\bar{{\mathtt{I}}}}$).}
\end{remark}


\subsection{Equilibrium under Full Information Disclosure}

Before analyzing the equilibria under Bayesian persuasion, we state the results obtained in our prior work \cite{satapathi2023coupled} for the setting where the agents are aware of their infection status, i.e., the FID setting with $\mu_{\St} = \mu_{\It} = 1$. In this case, $\pi^+[\St | \Stb] = 1$, $\pi^+[\It | \Itb] = 1$, $z_{\Itb} = z_{\It}$ and $z_{\Stb} = z_{\St}$. First we define a few quantities of interest:
\begin{multline*}
y^*_{\Ut} : = 1 - \frac{\gamma}{\betap}, \quad  \yint:= \frac{C_{\Pt}}{L(1-\alpha)\betap}, \quad y^*_{\Pt} := 1 - \frac{\gamma}{\alpha \betap}.
\end{multline*}
Next, we define the following population states that emerge as potential SNE:
\begin{align*}
\mathbf{E1} & = (y^*_{\Ut},1,0),  
\\ \mathbf{E2} & = (\yint,\zsint,0), \zsint = \frac{1}{1-\alpha}\Big[\frac{\gamma}{\betap(1-\yint)}-\alpha\Big],
\\ \mathbf{E3} & = (y^*_{\Pt},0,0).
\end{align*}

The following result from \cite{satapathi2023coupled} characterizes the SNE. 

\begin{proposition}[Equilibria under FID]\label{prop:equilibria-fid}
We have the following characterization of SNE when $\mu_{\St} = \mu_{\It} = 1$. 
\begin{enumerate}
\item $\mathbf{E1}$ is the SNE if and only if $y^*_{\Ut} < \yint$.
\item $\mathbf{E2}$ is the SNE if and only if $y^*_{\Pt} < \yint < y^*_{\Ut}$. 
\item $\mathbf{E3}$ is the SNE if and only if $y^*_{\Pt} > \yint$.
\end{enumerate}
\end{proposition}


\subsection{Equilibrium under Partial Information Disclosure}

We now characterize the SNE of this game under Assumption \ref{assumption:main}. \rev{Our main result shows that when $\yint$ is sufficiently small, then the SNE under PID coincides with the SNE under FID stated in Proposition \ref{prop:equilibria-fid}. However, for larger values of $\yint$, the infected proportion at the SNE can be made smaller by reducing $\mu_{\St}$ below $1$ till a certain threshold. If $\mu_{\St}$ is reduced further, then the infected proportion at the SNE can potentially be larger than the FID case.} We start with the following result \rev{which shows that if a non-zero fraction of individuals who receive signal $\Stb$ adopt protection, then all individuals who receive signal $\Itb$ adopt protection at a SNE}.


\begin{proposition}\label{prop:unstable}
    Let $(y^\star,z^\star_{\Stb},z^\star_{\Itb})$ be a SNE under Assumption \ref{assumption:main}. If $z^\star_{\Stb} < 1$, then $z^\star_{\Itb} = 0$. 
\end{proposition}
\begin{proof}
From \eqref{eq:del_Sbar}, it follows that 
$$ (1-\alpha) F[\St,\Ut] = - C_{\Pt} - \Delta U [\Stb]. $$
Substituting the above in \eqref{eq:del_Ibar}, we obtain
\begin{align*}
    \Delta U [\Itb] & = -\pi^+[\St | \Itb] (C_{\Ut} - \Delta U [\Stb] - C_{\Pt}) + C_{\Ut} - C_{\Pt}
    \\ & = (1-\pi^+[\St | \Itb])(C_{\Ut} - C_{\Pt}) + \pi^+[\St | \Itb] \Delta U [\Stb].
\end{align*}
If $z^\star_{\Stb} < 1$, then according to Definition \ref{def:SNE}, we must have $\Delta U [\Stb] \geq 0$. In addition, $C_{\Pt} < C_{\Ut}$ and $\pi^+[\St | \Itb] < 1$ for nonzero $y$. Consequently, we have $\Delta U [\Itb] > 0$ which implies $z^\star_{\Itb} = 0$. 
\end{proof}



Before proceeding further, we state the following quantities of interest, albeit with a slight abuse of notation to highlight the dependence on $\mu_{\St}$. Under the assumption $\mu_{\It} = 1$, we have $z_{\Itb} = z_{\It}$, and the effective infection rate
\begin{align}
    & \beta_{\mathtt{eff}}(z_{\Stb},z_{\Itb};\mu_{\St}) := (\beta_{\Pt} + (\beta_{\Ut} - \beta_{\Pt}) z_{\Itb}) \nonumber
    \\ & \qquad \qquad \times (\alpha + (1-\alpha)(z_{\Stb} \mu_{\St} + z_{\Itb} (1-\mu_{\St}))). \label{eq:betaeff_def}
\end{align}
Note that $\beta_{\mathtt{eff}}(z_{\Stb},z_{\Itb};\mu_{\St})$ is monotonically increasing in both $z_{\Itb}$ and $z_{\Stb}$, and is lower bounded by $\alpha \beta_{\Pt}$. Since it is assumed that $\gamma < \alpha \beta_{\Pt}$, there exists an endemic equilibrium 
\begin{equation}\label{eq:endemic_def}
    \yee(z_{\Stb},z_{\Itb};\mu_{\St}) = 1 - \frac{\gamma}{\beta_{\mathtt{eff}}(z_{\Stb},z_{\Itb};\mu_{\St})} \in (0,1),
\end{equation}
which is the stable equilibrium point for the epidemic dynamics \eqref{eq:sis_dynamics} for a given $(z_{\Itb},z_{\Stb})$ and $\mu_{\St}$. Since the R.H.S. of \eqref{eq:endemic_def} is monotonically increasing in $\beta_{\mathtt{eff}}(z_{\Stb},z_{\Itb};\mu_{\St})$, it follows that $\yee(z_{\Stb},z_{\Itb};\mu_{\St})$ is monotonically increasing in both its arguments for a given $\mu_{\St}$. We now define the following two functions:
\begin{align}
    h(z_{\Itb},\mu_{\St}) & := (1\!-\!\alpha) L (\beta_{\Pt} + (\beta_{\Ut} \!-\! \beta_{\Pt}) z_{\Itb})  \yee(1,z_{\Itb};\mu_{\St}) \!-\! C_{\Pt}, \nonumber
    \\ g(z_{\Itb},\mu_{\St}) & := \yee(1,z_{\Itb};\mu_{\St}) (C_{\Ut} - C_{\Pt}) \nonumber
\\ & \quad - (1\!-\!\mu_{\St})(1\!-\!\yee(1,z_{\Itb};\mu_{\St}))(\!-\!h(z_{\Itb},\mu_{\St})). \label{eq:function_g_zi_mus}
\end{align}

Note that when $z_{\Stb}=1$, $\beta_{\mathtt{eff}}(1,z_{\Itb};\mu_{\St})$ and thus, $\yee(1,z_{\Itb};\mu_{\St})$, are monotonically increasing in both $z_{\Itb}$ and $\mu_{\St}$ when the other variable is held constant. Thus, $h$ is monotonically increasing in both $z_{\Itb}$ and $\mu_{\St}$. Observe that when $h(z_{\Itb},\mu_{\St})\geq 0$, we have $g(z_{\Itb},\mu_{\St}) > 0$. If instead $h(z_{\Itb},\mu_{\St}) < 0$, then each of the terms in the product in the second line of \eqref{eq:function_g_zi_mus} is positive and decreasing in $\mu_{\St}$. As a result, $g(z_{\Itb},\mu_{\St})$ is monotonically increasing in $\mu_{\St}$ in the regime where $h(z_{\Itb},\mu_{\St}) < 0$. Similarly, it can be argued that $g(z_{\Itb},\mu_{\St})$ is monotonically increasing in $z_{\Itb}$ in this regime. 

The above properties have the following implications. If $g(0,0) \geq 0$, then $g(0,\mu_{\St}) \geq 0$ for all $\mu_{\St} \in [0,1]$. Otherwise, if $g(0,0) < 0$ and since $g(0,1)>0$, there exists a unique $\mu^{\star}_{\St} \in (0,1)$ for which $g(0,\mu^{\star}_{\St}) = 0$. We now define
\begin{align}\label{eq:def_mus_max}
    \mu^{\max}_{\St} := \begin{cases}
        & 0, \quad \text{if} \quad g(0,0) \geq 0,
        \\ & \mu^{\star}_{\St}, \quad \text{where } g(0,\mu^{\star}_{\St}) = 0, \mu^{\star}_{\St} \in (0,1). 
    \end{cases}
\end{align}
In particular, $\mu_{\St} \geq \mu^{\max}_{\St}\iff g(0,\mu_{\St}) \geq 0$. In addition, we also define the following quantities:
\begin{align}
    \zsdag & := \frac{\gamma - \alpha \betap (1 - \yint)}{\betap (1 - \alpha) (1 - \yint) \mu_{\St}}, \nonumber
    \\ \mu^{\min}_{\St} & := 1 - \frac{(\betau - \gamma) (C_{\Ut} - C_{\Pt})}{\gamma (C_{\Pt} - (1-\alpha) L (\beta_{\Ut} -\gamma))}.  \label{eq:def_mus_min}
\end{align}


\rev{We now state the main result of this paper. The proof is presented in the appendix.}

\begin{theorem}\label{theorem:main}
We have the following characterization of SNE $(y^\star,z^\star_{\Stb},z^\star_{\Itb})$ under Assumption \ref{assumption:main}.
\begin{enumerate}
\item $(y^*_{\Pt},0,0)$ is the SNE if and only if $y^*_{\Pt} > \yint$.
\item $(\yint, \zsdag, 0)$ is the SNE if and only if $\zsdag \in (0,1)$ or equivalently,
    \begin{align*}
        & 1 - \frac{\gamma}{\alpha \betap} < \yint < 1 - \frac{\gamma}{\betap (\alpha + (1 - \alpha) \mu_{\St})}.
    \end{align*}
\item $(\yee(1,0;\mu_{\St}),1,0)$ is the SNE if and only if 
    \begin{align*}
        &  \mu_{\St} \in [\mu^{\max}_{\St},1], \quad 1 - \frac{\gamma}{\betap (\alpha + (1 - \alpha) \mu_{\St})} \leq \yint. 
    \end{align*} 
\item     $(\yee(1,\zidag;\mu_{\St}),1,\zidag)$ with $\zidag \in (0,1)$ being the unique value satisfying $g(\zidag,\mu_{\St})=0$ is the SNE if and only if $\mu_{\St} < \mu^{\max}_{\St}$, and \rev{either of the following two conditions are satisfied:}
    \begin{align*}
        & \rev{L(1-\alpha)(\betau - \gamma) < C_{\Pt}} \quad \text{with} \quad \mu_{\St} >\mu^{\min}_{\St}, \quad \text{or} 
        \\ & \rev{L(1-\alpha)(\betau - \gamma) > C_{\Pt}}.
    \end{align*} 
\item $(1-\frac{\gamma}{\betau},1,1)$ is the SNE if and only if \rev{$L(1-\alpha)(\betau - \gamma) < C_{\Pt}$, and $\mu_{\St} < \mu^{\min}_{\St}$.}
\end{enumerate}
\end{theorem}


\subsection{Discussions}

We now discuss various implications of the above theorem. \rev{Note that the quantity $\yint$ is increasing in the parameter $\alpha$, and the normalized cost $\frac{C_{\Pt}}{L}$ which captures the relative cost of adopting protection compared to becoming infected. The first two cases of the theorem state that when $y^*_{\Pt} > \yint$, or $y^*_{\Pt} < \yint < \yee(1,0;\mu_{\St})$, the equilibrium under PID ($\mu_{\St}<1$) coincides with the equilibrium under FID ($\mu_{\St}=1$). In both these cases, the normalized cost is sufficiently small, and a nonzero fraction of agents receiving susceptible signal adopt protection.}



\begin{figure*}[t]
\centering
  \subfigure{\includegraphics[width=0.25\linewidth]{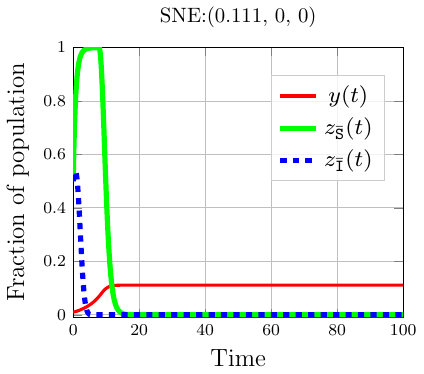}}
  ~~~~~~~~~~\subfigure{\includegraphics[width=0.25\linewidth]{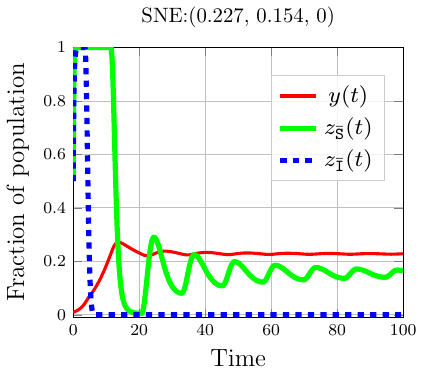}}
  ~~~~~~~~~~\subfigure{\includegraphics[width=0.25\linewidth]{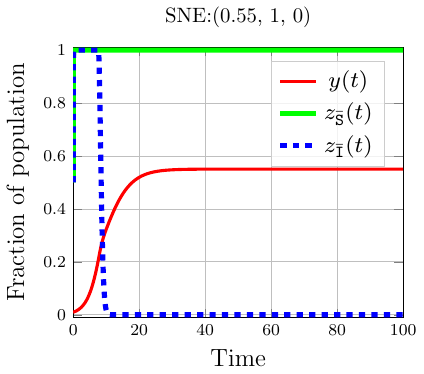}}
  \subfigure{\includegraphics[width=0.25\linewidth]{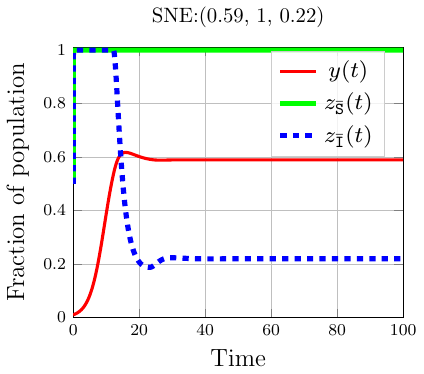}}
  ~~~~~~~~~~\subfigure{\includegraphics[width=0.25\linewidth]{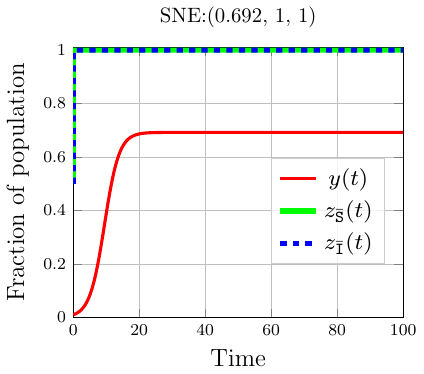}}
  ~~~~~~~~~~\subfigure{\includegraphics[width=0.24\linewidth]{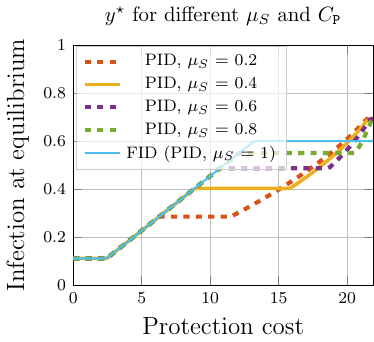}}
  \caption{Evolution and convergence of infected proportion, and fraction of population remaining unprotected upon receiving signals $\Stb$ and $\Itb$, to their respective stationary Nash equilibrium values $(y^\star,z^\star_{\Stb},z^\star_{\Itb})$ shown in the title, and (in bottom right panel) comparison of infection level at the stationary Nash equilibrium for different choice of $\mu_{\St}$ and $C_{\Pt}$.}
  \label{fig:Plot}
\end{figure*}


Thus, when $\mu_{\St}=1$, the infected proportion coincides with $\yint$ till $\yint$ increases to $1-\frac{\gamma}{\betap} = y^*_{\Ut}$. However, when $\mu_{S}<1$, the infected proportion grows with $\yint$ till the limit $\yee(1,0;\mu_{\St})$, which is smaller than $y^*_{\Ut}$. In other words, when $\yint$ grows beyond $\yee(1,0;\mu_{\St})$, the infected proportion grows with $\yint$ under FID, while it remains at $\yee(1,0;\mu_{\St})$ when $\mu_{\St} > \mu^{\max}_{\St}$ as stated in \rev{Case 3.} In this regime, the infected proportion under PID is smaller compared to the infection prevalence under FID. This is because while all the infected agents receive signal $\Itb$ and remain protected, a $1-\mu_{\St}$ fraction of susceptible agents also receive signal $\Itb$, and adopt protection. In contrast, under full information disclosure, this proportion does not necessarily adopt protection. Thus, a larger overall proportion of agents adopt protection leading to reduced infection prevalence at the SNE. 

However, this advantage does not hold if we reduce $\mu_{\St}$ below a threshold ($\mu^{\max}_{\St}$). \rev{Case 4 and Case 5 of Theorem \ref{theorem:main} show that when $\mu_{\St}$ is sufficiently small, then the agents receiving signal $\Itb$ are not sure regarding their true infection status as a large fraction of susceptible agents also receive this signal. Therefore, when $\frac{C_{\Pt}}{L}$ is sufficiently high, i.e., cost of protection is relatively large, all agents receiving susceptible signal and a nonzero fraction of agents receiving infected signal do not adopt protection. As a result, $y^\star$ starts to grow beyond $\yee(1,0;\mu_{\St})$, and potentially exceeds $\yint$ and eventually $y^*_{\Ut}$, as $\mu_{\St}$ continues to decline.}

\begin{remark}\label{remark:assumption}
\rev{The proposed framework can handle the case where $\mu_{\mathtt{I}} < 1$. However, when both $\mu_{\mathtt{S}} < 1$ and $\mu_{\mathtt{I}} < 1$, theoretical characterization becomes quite tricky and it would be difficult to obtain useful insights, compared to the relatively cleaner characterization derived in Theorem \ref{theorem:main} under $\mu_{\mathtt{I}} = 1$. In the following section, we numerically illustrate how the results vary when $\mu_{\mathtt{I}} < 1$. Similarly, if we assume $C_{\mathtt{P}} < C_{\mathtt{U}}$ instead of $C_{\mathtt{U}} < C_{\mathtt{P}}$, then in the FID case, only the magnitude of the infected proportions at the equilibria change, while the qualitative properties remain identical; see Remark 3 in \cite{satapathi2023coupled}. We expect a similar outcome in the present setting as well. For instance, it is easy to see that Proposition 2 would now change to: if $z^\star_{\bar{\mathtt{S}}} > 0$, then $z^\star_{\bar{\mathtt{I}}} = 1$, and, we would have a different set of equilibria in cases 2, 3, and 4 of Theorem \ref{theorem:main}. However, the qualitative properties of these new equilibria would be similar, without much new insights.}
\end{remark}


\section{Numerical Results}\label{sec:num_result}

In this section, we numerically examine the transient evolution of infected proportion and protection adoption behavior. In particular, we model the evolution of protection adoption decisions by the agents according to the \emph{Smith dynamics} \cite{sandholm2010population}, formally stated as
\begin{subequations}\label{eq:smith}
\begin{align}
\dot{z}_{\Stb} & = (1 \!-\! z_{\Stb}) \big[U[\Stb, \Ut] \!-\! U[\Stb, \Pt]\big]_{+} \!-\! z_{\Stb} \big[U[\Stb, \Pt] \!- U[\Stb, \Ut]\big]_{+}, \label{eq:smith_1}
\\ \dot{z}_{\Itb} & = (1 \!-\! z_{\Itb}) \big[U[\Itb, \Ut] \!-\! U[\Itb, \Pt]\big]_{+} \!-\! z_{\Itb} \big[U[\Itb, \Pt] \!- U[\Itb, \Ut]\big]_{+}, \label{eq:smith_2}
\end{align}
\end{subequations}
where operator $\big[a\big]_{+} := \max\{a, 0\}$. In particular, when $U[\Stb, \Pt] < U[\Stb, \Ut]$, the first term in the $\dot{z}_{\Stb}$ update equation is positive and the second term is $0$, leading to an increase in the proportion of unprotected agents; the converse holds when $U[\Stb, \Pt] > U[\Stb, \Ut]$. The update equation for $\dot{z}_{\Itb}$ is analogous. When $U[x, \Ut] - U[x, \Pt] = 0$ for $x\in\{\Stb,\Itb\}$, $\dot{z}_{x} = 0$, i.e., the proportions adopting protection remains unchanged. \rev{The Smith dynamics \cite{smith1984stability} is an instance of the class of {\it pairwise comparison} dynamics \cite{sandholm2015population}, and this class of dynamics possesses the Nash stationary property, i.e., the stationary points under the Smith dynamics corresponds to the Nash equilibrium (NE) of the underlying game, and vice versa. Therefore, agents can arrive at the NE by following this simple strategy update rule which corresponds to comparing the payoff among different available strategies. Furthermore, our result in Theorem \ref{theorem:main} is essentially a characterization of the stationary points of the coupled dynamics \eqref{eq:sis_dynamics} and \eqref{eq:smith}.}

We now show that state trajectories converge to the SNE characterized in Theorem \ref{theorem:main} for different parameter values. In particular, we set $\mu_{\St}=0.8$, vary the protection cost $C_{\Pt}$, and keep the remaining parameters unchanged with values stated in the following table.

\begin{table}[h]
\begin{center}
\begin{tabular}{|c | c | c | c | c | c | c | c | c |} 
 \hline
 $\alpha$ & $\beta_{\Pt}$ & $\beta_{\Ut}$ & $\gamma$ & $L$ & $C_{\Ut}$  & $y(0)$ & $z_{\Stb}(0)$ & $z_{\Itb}(0)$\\ [0.5ex] 
 \hline
 0.45 & 0.5 & 0.65 & 0.2 & 80 & 22 & 0.01 & 0.5 & 0.5\\ \hline
\end{tabular}
\end{center}
 \end{table}

\begin{figure*}[t]
\centering
  \subfigure{\includegraphics[width=0.25\linewidth]{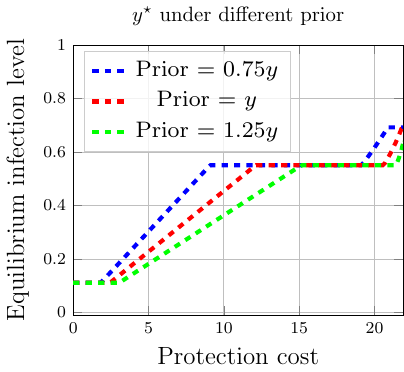}}
  ~~~~~~~~\subfigure{\includegraphics[width=0.25\linewidth]{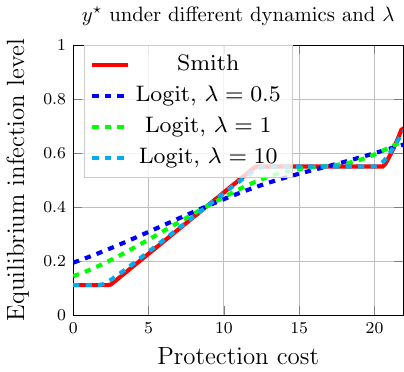}}
  ~~~~~~~~\subfigure{\includegraphics[width=0.3\linewidth]{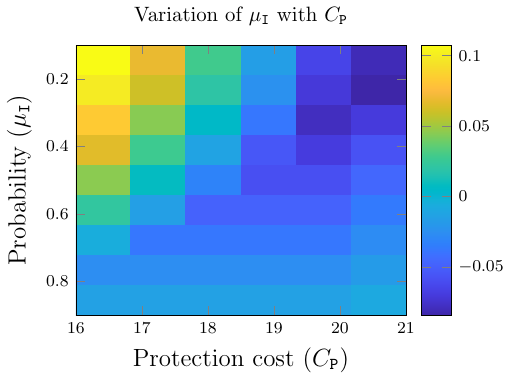}}
    \caption{\rev{Comparison of infection level at the stationary Nash equilibrium for different choice of $C_{\Pt}$ when (a) different prior information is used by the agents (left), (b) different values of bounded rationality parameter $\lambda$ (middle), and (c) difference in the infected proportion at the SNE when $\mu_{\It}=1$ and the infected proportion for other values of $\mu_{\It}$. All parameters are in accordance with the table, and $\mu_{\St}=0.8$.}}
  \label{fig:Plot_assumptions}
\end{figure*}

The trajectories of the fraction of population that is infected, and remains unprotected upon receiving signals $\Stb$ and $\Itb$ are shown in Figure \ref{fig:Plot}. The plots in the top row represent convergence to SNEs $(y^*_{\Pt}, 0, 0)$, $(\yint, \zsdag, 0)$ and $(\yee(1,0;\mu_{\St}),1,0)$, whereas, the plots in the left and middle panels of the second row depict convergence to $(\yee(1,\zidag;\mu_{\St}),1,\zidag)$ and $(1-\frac{\gamma}{\betau},1,1)$. The values of protection cost $C_{\Pt}$ that gives rise to SNE for each of the five sub-cases of Theorem \ref{theorem:main} are $1.5$, $5$, $15$, $21$, and $21.85$, respectively. The numerical values of the SNE are shown in the title of each plot. 

The bottom right plot captures infection level $y^\star$ at the SNE obtained by varying the protection cost $C_{\Pt}$ for different values of $\mu_{\St}$. Specifically, we let $C_{\Pt}$ vary from a very low value of $10^{-7}$ to $21.9$ (respecting the assumption $C_{\Pt} < C_{\Ut} = 22$). As $\mu_{\St}$ decreases below $1$, we observe that the infected proportion under PID coincides with the value under FID for the range $C_{\Pt}$ for which $\yint \leq \yee(1,0;\mu_{\St})$. When $\yint$ increases beyond $\yee(1,0;\mu_{\St})$, the equilibrium infection level under FID follows $\yint$ while it remains constant at $\yee(1,0;\mu_{\St})$ for PID. As $C_{\Pt}$ increases further, the threshold $\mu^{\max}_{\St}$ increases, and eventually exceeds our choice of $\mu_{\St}$. As a consequence, the infected proportion starts to increase, eventually exceeding the infection prevalence at FID and converging to $1-\frac{\gamma}{\betau}$ as $C_{\Pt}$ converges to $C_{\Ut}$. For a given $C_{\Pt}$ with $\yint \geq \yee(1,0;\mu_{\St})$, the infected proportion at the equilibrium decreases as $\mu_{\St}$ decreases from $1$ to $\mu^{\max}_{\St}$ beyond which the infection prevalence starts to climb up. 

Overall, this plot demonstrates that for certain parameter regimes, the infection level at equilibrium is smaller under an appropriate signalling scheme compared to the case when agents have accurate knowledge of their states, and vice versa. 

\rev{We now numerically investigate the consequence of relaxing some of our key assumptions. First, we examine the case where the prior probability of being infected is different from $y(t)$. The plot in the left panel of Figure \ref{fig:Plot_assumptions} shows the infected proportion at the SNE, $y^\star$, for different values of $C_{\Pt}$ when agents underestimate and overestimate the prior compared to $y(t)$. The results show that when agents overestimate the prior to be $1.25y(t)$, it leads to a smaller value of $y^\star$, and vice versa when the prior is assumed to be $0.75y(t)$. Nevertheless, the variation of $y^\star$ as a function of $C_{\mathtt{P}}$ is qualitatively similar when the prior information used by the agents is larger or smaller compared to $y(t)$.} 

\rev{The discussion thus far has focused on the Nash equilibrium (NE) of the proposed setting as NE remains the most prominent and widely accepted solution concept in game theory. Nevertheless, human decision making is often subject to different behavioral biases. In the middle panel of Figure \ref{fig:Plot_assumptions}, we illustrate the impact of bounded rationality on the infected proportion at the SNE for different values of $C_{\Pt}$ by assuming that agents update their strategies following the {\it logit} dynamics \cite{sandholm2010population}. The parameter $\lambda$ present in the logit dynamics captures the degree of bounded rationality; if $\lambda$ is large, then agents choose the action that gives larger payoff with a greater probability, and vice versa. The figure shows that the outcome under $\lambda=10$ nearly coincides with the SNE, while it can be somewhat different for smaller values of $\lambda$. In particular, there is a range of $C_{\Pt}$ over which bounded rationality leads to a smaller proportion of infected agents at the SNE.}

\rev{Finally, we examine the outcome when $\mu_{\mathtt{I}} \in (0,1)$. The heatmap on the right panel of Figure \ref{fig:Plot_assumptions} shows the difference between the infected proportion at the SNE when $\mu_{\mathtt{I}} = 1$ and the infected proportion at the SNE for other values of $\mu_{\mathtt{I}}$. The results show that when $\mu_{\mathtt{I}}$ is close to $1$, i.e., larger than around $0.7$, then the corresponding infected proportion at the equilibrium is larger compared to the case when $\mu_{\mathtt{I}}=1$. However, when $\mu_{\mathtt{I}}$ is close to $0$, which is an extreme case, and the cost of protection is small, the corresponding infected proportion at the equilibrium is smaller compared to when $\mu_{\mathtt{I}}=1$. }


\section{Conclusion and Future Work}

In this work, we examined the effectiveness of Bayesian persuasion in containing the prevalence of SIS epidemic when the agents are not aware of their true infection status while making decisions. Specifically, we identify conditions under which partial information disclosure to susceptible agents reduces the fraction of infected agents at the stationary Nash equilibrium. We further showed that when the signal revealed to susceptible agents is not sufficiently close to the truth, then it may lead to infected agents not adopting protection, and the fraction of infected agents at the equilibrium could be larger compared to the FID regime. 

Our work is one of the first to rigorously investigate the impact of Bayesian persuasion in containing spreading processes. There remain several interesting directions for future research. \rev{Extending the present work to the setting where the agents and/or the sender are non-myopic or forward-looking, potentially leveraging recent works \cite{gan2022bayesian,wu2022sequential,lehrer2021markovian}, remains an important open problem. Similarly, analyzing stability of the coupled dynamics \eqref{eq:sis_dynamics} and \eqref{eq:smith}, as well as designing feedback schemes to dynamically modulate the signalling scheme to steer the strategies of the agents to a desirable equilibrium point (potentially building upon \cite{martins2023epidemic}) is a challenging open problem. In addition, extending the present setting to the case where the agents are heterogeneous (e.g., with respect to their degrees), and to other classes of epidemics models (such as the susceptible-infected-recovered-susceptible model), and protection schemes (e.g., vaccination) remain as promising directions for future research.}


\section*{Acknowledgement}
We thank Prof. Ankur Kulkarni (IIT Bombay) for his suggestions. 


\appendix[Proof of Theorem \ref{theorem:main}]

\begin{proof} We prove each of the statements in the theorem in the sequence they are stated. 

\noindent {\bf Case 1:} $(y^*_{\Pt},0,0)$. First observe that $\yee(0,0;\mu_{\St}) = y^*_{\Pt}$. Note that the necessary condition for $z^\star_{\Stb} = 0$ is $\Delta U[\Stb] \geq 0$, or equivalently, 
\begin{align*}
    & (1-\alpha) L (\beta_{\Pt} + (\beta_{\Ut} - \beta_{\Pt}) z^\star_{\Itb}) y^*_{\Pt}- C_{\Pt} \geq 0
    \\ \iff & (1-\alpha) L (\beta_{\Pt}) y^*_{\Pt}- C_{\Pt} \geq 0
    \\ \iff & y^*_{\Pt} \geq \frac{C_{\Pt}}{(1-\alpha) L (\beta_{\Pt})} = \yint. 
\end{align*}
For sufficiency, assume that $y^*_{\Pt} > \yint$. Let $z^\star_{\Stb}$ and $z^\star_{\Itb}$ be the proportions of individuals who remain unprotected upon receiving signal $\Stb$ and $\Itb$, respectively. Then, $\yee(z^\star_{\Stb},z^\star_{\Itb};\mu_{\St}) \geq \yee(0,0;\mu_{\St})$ since $\yee(z_{\Stb},z_{\Itb};\mu_{\St})$ is monotonically increasing in both $z_{\Stb}$ and $z_{\Itb}$. We now compute 
\begin{align*}
\Delta U[\Stb] & = (1-\alpha) L (\beta_{\Pt} + (\beta_{\Ut} - \beta_{\Pt}) z^\star_{\Itb}) \yee(z^\star_{\Stb},z^\star_{\Itb};\mu_{\St})- C_{\Pt}
\\ & \geq (1-\alpha) L (\beta_{\Pt}) \yee(0,0;\mu_{\St})- C_{\Pt} > 0,
\end{align*}
under our hypothesis, and $z^\star_{\Itb} = 0$ follows from Proposition \ref{prop:unstable}.


\noindent {\bf Case 2:} $(\yint, \zsdag, 0)$. Observe that $\zsdag > 0$ is equivalent to
\begin{align*}
    & \frac{\gamma - \alpha \betap (1 - \yint)}{\betap (1 - \alpha) (1 - \yint) \mu_{\St}} > 0
    \\ \iff & \gamma > \alpha \betap (1 - \yint) \iff 1 - \frac{\gamma}{\alpha \betap} < \yint. 
\end{align*}
Similarly, $\zsdag < 1$ is equivalent to
\begin{align*}
    & \gamma - \alpha \betap (1 - \yint)< \betap (1 - \alpha) (1 - \yint) \mu_{\St}
    \\ \iff & \gamma < \betap (1 - \yint) (\alpha + (1 - \alpha) \mu_{\St})
    \\ \iff & \yint < 1 - \frac{\gamma}{\betap (\alpha + (1 - \alpha) \mu_{\St})}.
\end{align*}

Since $\zsdag \in (0,1)$, it follows from Proposition \ref{prop:unstable} that $\zibar = 0$. The endemic equilibrium is given by 
\begin{align*}
    \yee(\zsdag,0;\mu_{\St}) & = 1-\frac{\gamma}{\betap (\alpha+(1-\alpha)(\zsdag \mu_{\St}))}
    \\ & = 1-\frac{\gamma}{\betap \big(\alpha+(\frac{\gamma - \alpha \betap (1 - \yint)}{\betap (1 - \yint)})\big)}
    \\ & = 1-\frac{\gamma \betap (1 - \yint) }{\betap \gamma} = \yint. 
\end{align*}

The necessary condition pertaining to $\zsdag \in (0, 1)$ is $\Delta U[\Stb] = 0$. Thus, we compute $\Delta U[\Stb]$ as
\begin{align*}
    & (1-\alpha) L (\beta_{\Pt} + (\beta_{\Ut} - \beta_{\Pt}) \zibar) \yee(\zsdag,0;\mu_{\St})- C_{\Pt}
    \\ = ~ & (1-\alpha) L \beta_{\Pt} \yint - C_{\Pt} = 0. 
\end{align*}

For sufficiency, suppose the hypothesis stated in the theorem holds true. It follows from \eqref{eq:del_Sbar} that the function
$$
\Delta U[\Stb](z_{\Stb},z_{\Itb}) = (1-\alpha) L (\beta_{\Pt} + (\beta_{\Ut} - \beta_{\Pt}) z_{\Itb}) \yee(z_{\Stb},z_{\Itb};\mu_{\St})- C_{\Pt}
$$
is monotonically increasing in both of its arguments. Under our hypothesis, we have
\begin{align*}
\Delta U[\Stb](0,0) & = (1-\alpha) L \beta_{\Pt} \yee(0,0;\mu_{\Stb})- C_{\Pt}
\\ & = (1-\alpha) L \beta_{\Pt} \big[ 1 - \frac{\gamma}{\alpha \betap} \big] - C_{\Pt} < 0
\end{align*}
which implies $\zsbar \neq 0$. Similarly, we compute
\begin{align*}
& \Delta U[\Stb](1,0) = (1-\alpha) L \beta_{\Pt} \yee(1,0;\mu_{\Stb})- C_{\Pt}
\\ & \quad = (1-\alpha) L \beta_{\Pt} \big[ 1-\frac{\gamma}{\betap (\alpha+(1-\alpha)(\mu_{\St}))} \big]- C_{\Pt}
>0.
\end{align*}
As a result, $\Delta U[\Stb](1,z_{\Itb}) > 0$ for all $z_{\Itb} \in [0,1]$. Therefore, $\zsbar \neq 1$. Nevertheless, due to the monotonicity of $\Delta U[\Stb](z_{\Stb},0)$, there exists a unique $\zsbar \in (0,1)$ at which $\Delta U[\Stb](\zsbar,0) = 0$. Following above calculations, it can be easily shown that $\zsbar = \zsdag$ and $\yee(\zsdag,0;\mu_{\St}) = \yint$.


\noindent {\bf Case 3:} $(\yee(1,0;\mu_{\St}),1,0)$. We start by establishing the necessary conditions, which for $\zsbar=1$ is
    \begin{align*}
        & \Delta U[\Stb](\yee(1,0;\mu_{\St}),1,0) \leq 0
        \\ \iff & (1-\alpha) L \betap \yee(1,0;\mu_{\St}) \leq C_{\Pt}
        \\ \iff & \yee(1,0;\mu_{\St}) = 1-\frac{\gamma}{\betap(\alpha+(1-\alpha)\mu_{\St})} \leq \yint. 
    \end{align*} 

    In order to compute the difference in utilities at the SNE for an agent that receives signal $\Itb$, we first define
    \begin{align*}
    & \pi^{+}[\St|\Itb](1,z_{\Itb};\mu_{\St}) = 1 - \frac{\yee(1,z_{\Itb};\mu_{\St})}{1-\mu_{\St}+\mu_{\St}\yee(1,z_{\Itb};\mu_{\St})}.
    \end{align*}
    We now compute
    \begin{align*}
    & \Delta U[\Itb](\yee(1,z_{\Itb};\mu_{\St}),1,z_{\Itb}) 
    \\ & \qquad = (1-\pi^+[\St | \Itb](1,z_{\Itb};\mu_{\St}))(C_{\Ut} - C_{\Pt}) \nonumber
    \\ & \qquad \qquad + \pi^+[\St | \Itb](1,z_{\Itb};\mu_{\St}) \Delta U [\Stb](\yee(1,z_{\Itb};\mu_{\St}),1,z_{\Itb})
    \\ \Rightarrow & (1-\mu_{\St}+\mu_{\St}\yee(1,z_{\Itb};\mu_{\St})) \Delta U[\Itb](\yee(1,z_{\Itb};\mu_{\St}),1,z_{\Itb})
    \\ & \qquad = g(z_{\Itb},\mu_{\St}),
    \end{align*} 
    where the function $g$ is defined in \eqref{eq:function_g_zi_mus}. In particular, when $z_{\Stb}=1$, the sign of $\Delta U[\Itb]$ coincides with the sign of the function $g$. 
    
    We now state the necessary condition for $\zibar=0$ as
    \begin{align*}
        \Delta U[\Itb](\yee(1,0;\mu_{\St}),1,0) \geq 0 & \iff g(0,\mu_{\St}) \geq 0 
        \\ & \iff \mu_{\St} \in [\mu^{\max}_{\St},1],
    \end{align*}
    following the definition of $\mu^{\max}_{\St}$ in \eqref{eq:def_mus_max}. 
     
    The necessary conditions are also sufficient in this case. Specifically, for $\mu_{\St} > \mu^{\max}_{\St}$, we have $g(0,\mu_{\St}) \geq 0$. From the monotonicity of $g$ in $z_{\Itb}$, it follows that $g(z_{\Itb},\mu_{\St}) \geq 0$ for all $z_{\Itb} \in [0,1]$, or equivalently, 
    $$ \Delta U[\Itb](\yee(1,z_{\Itb};\mu_{\St}),1,z_{\Itb}) \geq 0$$
    for every $z_{\Itb} \in (0,1]$. As a result, any $z^\star_{\Itb} \neq 0$ cannot be present at a SNE. When $z_{\Itb}=0$, we have 
    $$\Delta U[\Stb](\yee(z_{\Stb},0;\mu_{\St}),z_{\Stb},0) < \Delta U[\Stb](\yee(1,0;\mu_{\St}),1,0) \leq 0$$ 
    for any $z_{\Stb} \in [0,1)$ due to the monotonicity of $\Delta U[\Stb]$ in $z_{\Stb}$; the second inequality above follows from our hypothesis on $\yint$. As a result, $z^\star_{\Stb} \neq 1$ cannot be present at a SNE. 
    Finally, at $z^\star_{\Stb} = 1$ and $z^\star_{\Itb} = 0$, the sufficient conditions for SNE are  satisfied under our hypothesis as shown above. 
    

\noindent {\bf Case 4:} $(\yee(1,\zidag;\mu_{\St}),1,\zidag)$. We first show the sufficiency part. Recall that when $\mu^{\max}_{\St} > 0$, then $g(0,\mu^{\max}_{\St})=0$. From the monotonicity of $g$, it follows that $g(0,\mu_{\St}) < 0$ for all $\mu_{\St} < \mu^{\max}_{\St}$. Thus, it follows from similar arguments as the previous case that $\Delta U[\Itb](\yee(1,0;\mu_{\St}),1,0) < 0$ for $\mu_{\St} < \mu^{\max}_{\St}$. Therefore, $\zibar=0$ can not arise at the SNE. We now observe that $\yee(1,1;\mu_{\St}) = 1 - \frac{\gamma}{\betau}$ or equivalently, $\betau (\yee(1,1;\mu_{\St})) = \betau - \gamma$.

When $\mu_{\St} > \mu^{\min}_{\St}$ \rev{and $C_{\Pt} > (1-\alpha) L (\beta_{\Ut} -\gamma)$}, we obtain
\begin{subequations}\label{eq:case4_theorem}
  \begin{align}
& 1 - \mu_{\St} < 1- \mu^{\min}_{\St} = \frac{(\betau - \gamma) (C_{\Ut} - C_{\Pt})}{\gamma (C_{\Pt} - (1-\alpha) L (\beta_{\Ut} -\gamma))}
\\ \iff & (C_{\Ut} - C_{\Pt}) \betau (\yee(1,1;\mu_{\St})) \nonumber
\\ & \quad > (1 - \mu_{\St}) \gamma (C_{\Pt} - (1-\alpha) L (\beta_{\Ut} -\gamma))
\\ \iff & (C_{\Ut} - C_{\Pt}) \yee(1,1;\mu_{\St}) \nonumber
\\ & > (1 \!-\! \mu_{\St}) (1\!-\!\yee(1,1;\mu_{\St})) (\!-\!\Delta U[\Stb](\yee(1,1;\mu_{\St}),1,1)) \nonumber
\\ \iff & g(1,\mu_{\St}) > 0,
\end{align}  
\end{subequations}
which implies that $\zibar=1$ can not arise at the SNE. \rev{The same conclusion holds when $C_{\Pt} < (1-\alpha) L (\beta_{\Ut} -\gamma)$ which implies that $\mu_{\St} < 1 < \mu^{\min}_{\St}$.}

Nevertheless, since $g(z_{\Itb},\mu_{\St})$ is monotonically increasing in $z_{\Itb}$, there exists a unique $\zidag \in (0,1)$ at which $$g(\zidag,\mu_{\St})=0 \iff \Delta U[\Itb](\yee(1,\zidag;\mu_{\St}),1,\zidag) = 0.$$ Consequently, $$\Delta U[\Stb](\yee(1,\zidag;\mu_{\St}),1,\zidag) < 0 \implies \zsbar=1.$$ 

We next show the necessity. Since $\zidag \in (0,1)$, we must have $\Delta U[\Itb](\yee(1,\zidag;\mu_{\St}),1,\zidag) = 0$ or equivalently, $g(\zidag,\mu_{\St})=0$. Due to monotonicity of $g$ in the first argument, we obtain
\begin{align*}
& g(0,\mu_{\St})<0 \iff \mu^{\max}_{\St}>0, \mu_{\St}< \mu^{\max}_{\St}.
\end{align*}
Finally, $g(\zidag,\mu_{\St})=0$ implies $g(1,\mu_{\St})>0$ which is equivalent to the conditions stated in the statement of the Theorem following \eqref{eq:case4_theorem}. 

\noindent {\bf Case 5:} $(1-\frac{\gamma}{\betau},1,1)$. We start by establishing the necessary condition. From Proposition \ref{prop:unstable}, it follows that $z^\star_{\Itb} = 1 \implies z^\star_{\Stb} = 1$. In addition, $\yee(1,1;\mu_{\St}) = 1-\frac{\gamma}{\betau}$ which is independent of $\mu_{\St}$. The necessary condition for $z^\star_{\Itb} = 1$ is $$\Delta U[\Itb](\yee(1,1;\mu_{\St}),1,1) \leq 0 \iff g(1,\mu_{\St})\leq 0.$$ In addition, $\Delta U [\Stb](\yee(1,1;\mu_{\St}),1,1) < 0$ yields
\begin{align*}
    & (1-\alpha) L \beta_{\Ut} \yee(1,1;\mu_{\St})- C_{\Pt} < 0
    \\ \implies & L(1-\alpha)(\betau - \gamma) < C_{\Pt}.
\end{align*} 
It follows from \eqref{eq:case4_theorem} that when $L(1-\alpha)(\betau - \gamma) < C_{\Pt}$ and $g(1,\mu_{\St})\leq 0$, we must have $\mu_{\St} < \mu^{\min}_{\St}$.


For sufficiency, suppose the hypotheses stated in the theorem hold true. Let $(\yee(z^\star_{\Stb},z^\star_{\Itb};\mu_{\St}), z^\star_{\Stb},z^\star_{\Itb})$ be a SNE. We compute 
\begin{align*}
\Delta U[\Stb] & = (1-\alpha) L (\beta_{\Pt} + (\beta_{\Ut} - \beta_{\Pt}) z^\star_{\Itb}) \yee(z^\star_{\Stb},z^\star_{\Itb};\mu_{\St})- C_{\Pt}
\\ & \leq (1-\alpha) L (\beta_{\Ut}) \yee(1,1;\mu_{\St})- C_{\Pt} < 0,
\end{align*}
where the first inequality follows from the monotonicity of $\yee(z^\star_{\Stb},z^\star_{\Itb};\mu_{\St})$ in both of its arguments for a given $\mu_{\St}$, and the second inequality holds due to our hypothesis. Consequently, we must have $z^\star_{\Stb} = 1$. 

It follows from \eqref{eq:case4_theorem} that $\mu_{\St} < \mu^{\min}_{\St}$ together with $C_{\Pt} > (1-\alpha) L (\beta_{\Ut} -\gamma))$ implies 
\begin{align*}
& g(1,\mu_{\St})\leq 0 \iff \Delta U[\Itb](\yee(1,1;\mu_{\St}),1,1) \leq 0
\\ & \implies \Delta U[\Itb](\yee(1,z^\star_{\Itb};\mu_{\St}),1,z^\star_{\Itb}) \leq 0 \quad \forall z^\star_{\Itb} \in [0,1],
\end{align*}
due to the monotonicity of $g$ in $z^\star_{\Itb}$. Therefore, we must have $z^\star_{\Itb}=1$. This concludes the proof.
\end{proof}

 
\bibliographystyle{IEEEtran}
\bibliography{ref}

\end{document}